\documentstyle[12pt,aaspp4,psfig]{article} 

\begin{document}

\title{RXTE Observations of the Anomalous Pulsar 4U 0142+61}
\author
{Colleen~A.~Wilson, Stefan Dieters\altaffilmark{1}, Mark H.
Finger\altaffilmark{2}, D. Matthew Scott\altaffilmark{2}, Jan van
Paradijs\altaffilmark{1}}
\affil{\footnotesize ES 84 Space Sciences Laboratory, NASA/Marshall
Space Flight Center, Huntsville, AL 35812;
colleen.wilson@msfc.nasa.gov, stefan.dieters@msfc.nasa.gov, finger@gibson.msfc.nasa.gov,
scott@gibson.msfc.nasa.gov, jvp@astro.uva.nl}
\altaffiltext{1}{University of Alabama in Huntsville}
\altaffiltext{2}{Universities Space Research Association}

\begin{abstract}

We observed the anomalous X-ray pulsar 4U 0142+61 using the Proportional Counter Array
(PCA) aboard the {\em Rossi X-ray Timing Explorer (RXTE)} in March 1996. 
The pulse frequency was measured as $\nu$ = 0.11510039(3) Hz with an upper 
limit of $|\dot \nu| \leq 4 \times 10^{-13}$~Hz~s$^{-1}$ upon the short term 
change in frequency over the 4.6 day span of the observations.  A compilation 
of all historical measurements showed an overall spin-down trend with slope 
$\dot \nu = -3.0 \pm 0.1 \times 10^{-14}$~Hz~s$^{-1}$.  Searches for orbital 
modulations in pulse arrival times yielded an upper limit of 
$a_{\rm x} \sin i \lesssim $ 0.26~lt-s (99\%
confidence) for the period range 70~s to 2.5 days. These limits combined with 
previous optical limits and evolutionary arguments suggest that 4U 0142+61 is 
probably not a member of a binary system.

\end{abstract}


\section{Introduction}             
  
Most known accreting X-ray pulsars are members of high-mass X-ray binaries
(HMXBs) containing an OB companion (\cite{vanPara95}; \cite{Bildsten97}).  Comparatively, the 
number of pulsars found in low-mass X-ray binaries (LMXBs) is quite small.  
Only five pulsars are known to be members of LMXBs: Her X--1, GX 1+4, GRO
J1744--28, 4U 1626--67, and SAX J1808.4-3658 = XTE J1808-369
(\cite{Chakrabarty98}), which have known orbital periods or companions 
(\cite{Bildsten97} and references therein.) In addition to the five known LMXB
pulsars, there are 6 to 8 pulsars which are not members of HMXBs: 4U 0142+61, 
1E 1048.1--5937, 1E 2259+586, RX J1838.4--0301 (see \cite{Mereghetti97} for an
alternative interpretation of this system), 1E 1841--045, 
1RXS J170849.0--400910 (\cite{Vasisht97} and references therein), AX
J1845--0258 (\cite{Gotthelf98}), and possibly RX J0720.4--3125 
(\cite{Haberl97}) . These pulsars, called 
anomalous or braking X-ray pulsars (\cite{Vanparadijs95}; \cite{Mereghetti95}; 
\cite{Ghosh97}), all have spin periods in the 5-11~s range, which is extremely
narrow compared to the 69~ms-2.8~hr range of spin periods of HMXB pulsars.
No optical counterparts of these systems have been detected to date, and 
4U 0142+61 (\cite{Steinle87}; \cite{White87}; \cite{Coe98}), 1E 1048.1--5937
(\cite{Mereghetti92}), 1E 2259+586, RX J1838.4--0301 (\cite{Coe98} and
references therein), RX J0720.4--3125 
(\cite{Haberl97}; \cite{Motch98}) have optical limits that rule out a 
high-mass companion.
These pulsars are also characterized by the following properties: (1) a 
general spin down trend with time scales $P/|\dot P| = 10^3-10^5$ years, while 
HMXBs with similar periods show episodes of spin-up and spin-down
(\cite{Bildsten97}); (2) No evidence of orbital
periodicity to date; (3) a low X-ray luminosity of $10^{34}-10^{36}$ ergs s$^{-1}$ that is quite constant on time scales of days to $\sim 10$ years; (4) a
very soft X-ray spectrum usually well described by a combination of a blackbody
with effective temperature $\sim$ 0.3-0.4 keV and a photon power law 
($f(E) = A E^{-\gamma}$) with photon index $\gamma = 3-4$ 
(\cite{Mereghetti98b}); and (5) a relatively young age ($\lesssim 10^5$~yr) as
inferred from the low
galactic scale height ($\sim 100$ pc r.m.s.) of the anomalous X-ray pulsars 
(AXPs) as a group (\cite{Vanparadijs95}) and the apparent association of 
1E 2259+587 (\cite{Fahlman81}), 1E 1841--045 (\cite{Vasisht97}), and RX 
J1838.4--0301 (\cite{Schwentker94}) with supernova remnants. 

The persistent X-ray source 4U 0142+61 was discovered with {\em Uhuru}
(\cite{Forman78}). Pulsations at 8.7~s were discovered in 1984 August 
{\em EXOSAT} data (\cite{Israel94}).  A 25 minute modulation was also seen in 
these data (\cite{White87}), but was later found to be due to the
nearby X-ray transient pulsar RX J1046.9+6121 (\cite{Motch91}; \cite{Hellier94}).  
The {\em ASCA} 2-10 keV spectrum of 4U 0142+61 was well described by a blackbody with effective
temperature $0.386 \pm 0.005$ keV plus a power law with photon index $3.67 \pm
0.09$ and an absorption column $N_{\rm H} = 9.5 \pm 0.4 \times 10^{21}$ 
cm$^{-2}$ (\cite{White96}). This spectrum was consistent with an earlier 
spectrum measured with {\em EXOSAT} when RX J0146.9+6121 was not present
(\cite{White87}).
The spin frequency of 4U0142+61 was measured with {\em Einstein} (\cite{White96}),
{\em EXOSAT} (\cite{Israel94}), {\em ROSAT} (\cite{Motch91}; \cite{Hellier94}), 
{\em ASCA} (\cite{White96}), and {\em RXTE} (this paper). The spin frequency 
history is shown in Figure~\ref{fig:period}.  All high significance historical 
measurements of the spin frequency follow the general spin-down trend of 
$\dot \nu = -3.1 \pm 0.1\times 10^{-14}$ Hz s$^{-1}$, which corresponds to a
spin-down timescale $\nu/|\dot \nu| \approx$ 123,000 yr.  Three frequency 
measurements from 1985 EXOSAT post discovery data and 1991 ROSAT data 
(\cite{Israel94}), which Israel et al. (1994) found to have high probabilities
of chance occurrence, have been omitted from Figure~\ref{fig:period}. 

Israel et al. (1994) \nocite{Israel94} performed an orbital period search on 
$\sim 12$~hr of {\em EXOSAT} ME data from 1984 August 27-28.  The search was 
carried out for periods from 430~s to 43,000~s, assuming a circular orbit. No 
significant orbital signature was found. Other {\em EXOSAT} measurements in 
1985 November and December did not have sufficient time resolution to 
unambiguously detect 4U 0142+61.

Here we present results from {\em RXTE} observations of 4U 0142+61, obtained in
1996 March. The observation times and durations are listed in Table~1. 
An orbital period search is performed for periods between $\sim$ 600~s and 4.6 
days. Power spectra are presented and compared to white noise 
spectra with identical windowing.  Upper limits are placed on the size of the 
allowed orbits for different trial periods. Next a modified algorithm is used 
to search for orbital periods shorter than $\sim$ 600~s and to place upper 
limits on the size of allowed orbits. {\em RXTE} results are compared to previous results
obtained with {\em EXOSAT}. Allowed companion types based upon these orbital
limits and previous optical limits are discussed.
       
\section{Observations and Analyses}

The Proportional Counter Array (PCA) on {\em RXTE} consists of five
xenon/methane multi-anode proportional
counters sensitive to photons from 2-60 keV and has a total collecting area of
6500 cm$^2$. The PCA is a collimated instrument with an approximately circular
field-of-view with a FWHM of about 1\arcdeg\ (\cite{Jahoda96}).  PCA
observations of 4U 0142+61 were performed on 1996 March 25 and 28-30. Details
of these observations are given in Table~1. The
field of view also contained the Be/X-ray pulsar RX J0146.9+6121
(\cite{Motch91}; \cite{Hellier94}; \cite{Haberl98}).  For 1996 March 28-30, Good Xenon data 
(time and detector tagged events with 1~$\mu$s time resolution
and full energy resolution) were used for our analysis.  For 1996 March 25, 
Event Mode data, which have lower time resolution and similar energy resolution,
were used. To maximize signal to noise ratio, data from only the top xenon layer of the PCA  and from the energy
range 3.7-9.2 keV were selected. The light curves were binned to 125~ms time 
resolution and the times were corrected to the solar system barycenter.  

The pulse frequency of 4U 0142+61 was determined to be $\nu_{\rm spin} = 
0.11510051(8)$ by an epoch-folded search of data from 1996 March 25 and 28-30 for
the frequency range 0.1139 - 0.1165 Hz.
Assuming the pulse frequency remained constant at this value across our 
observations, we divided the data into segments and determined phase offsets 
and intensities for each segment. A 
template profile was created by epoch-folding the first 80,000 seconds of data
from March 28-30. The template profile is shown in Figure~\ref{fig:profile}. 
The data were then divided into 250 second segments. In each segment the data 
were epoch-folded at the pulse frequency. The template and the pulse profiles from
each segment were represented by a Fourier expansion in pulse phase.  The 
profiles were limited to the first 3 Fourier coefficients, which contained 
98.6\% of the power in the template. These Fourier coefficients, $a_k$, were 
calculated as 
\begin{equation}
  a_k = \frac{1}{N} \sum^N_{j=1} R_j e^{-i 2 \pi j k/N}
\end{equation}  
where $k$ is the harmonic number, $i=(-1)^{1/2}$, $N=16$ is the number of phase bins
in the pulse profile, and $R_j$ is the count rate in the $j$th bin of the pulse
profile.  The pulse phase offset of each segment with respect to the template
was calculated by cross-correlating each phase fold with the template.  The
cross-correlation is equivalent to a fit to 
\begin{equation}
 \chi^2 = \sum^3_{k=1} \frac{|a_k-I e^{i 2 \pi k \phi}
t_k|^2}{\sigma_{a_k}^2}
\end{equation}
where $t_k$ is the Fourier coefficient of harmonic number $k$ of the template
profile, $I$ is the relative intensity of the measured profile with respect to
the template, and $\phi$ is the phase offset of the measured profile relative to
the template profile. The errors on the phase offsets were weighted according 
to the amount of data in each segment. Figure~\ref{fig:phases} shows the phase
offsets with a best fit quadratic removed. A total of 7 points, which deviated 
from this fit by more than 0.25 cycles, were discarded. For 5 of these
outliers, the 250 s bins were less than $1/3$ full. The remaining two outliers,
which both contained less than 200 s of data and were offset by about $1/2$
cycle, contained double-peaked pulse profiles. A quadratic fit to the
remaining points yielded a best fit frequency of $\nu = 0.11510039(3)$~Hz at MJD
50169.835 (consistent with our earlier epoch-folding results) and a 3~$\sigma$
upper limit of $\dot \nu \lesssim 4 \times 10^{-13}$ Hz s$^{-1}$ with 
$\chi^2/182 = 1.05$. 

The phase offsets were searched for an orbital periodicity using a
Lomb-Scargle periodogram (\cite{Press92} and references therein) shown in
Figure~\ref{fig:lomb}.  The top panel is the Lomb-Scargle periodogram for 
the entire 4 day interval.  The largest peak at $\approx$ 586~s has a  
probability of chance coincidence of 65\%. The center panel is the Lomb-Scargle periodogram for
March 28-30 only.  The largest peak at  $\approx$ 1680~s has a probability of 
chance coincidence of 78\%. These probabilities were computed using Monte-Carlo
simulations of 10,000 white noise realizations. The phase offsets were simulated as white noise 
with zero mean and standard deviations equal to the measurement errors on the 
phase offsets along with the actual time tags corresponding to the phase 
offsets. The cumulative probability distribution of the peak noise power was 
generated by retaining the power in the largest noise peak for each trial.
The bottom panel is the Lomb-Scargle periodogram for white noise with mean and
variance equal to that of the phase offsets with time tags corresponding to the
4 day interval. From the noise power, it is evident that much of the higher
frequency variability in the power spectra is due to the windowing of the data.
The Lomb-Scargle technique removes a constant from the data. Any variations in
count rate that deviate from a constant result in low frequency variations in 
the power spectrum. This analysis revealed no significant orbital signatures. 

Previous upper limits, $a_{\rm x} \sin i \lesssim 0.37$ lt-s (430~s $\lesssim
P_{\rm orb} \lesssim 43000$~s) estimated from {\em EXOSAT} data by Israel et al.
(1994), \nocite{Israel94} were calculated using the method of Van der Klis
(1989) \nocite{vanderklis89}. This method assumes that the total power in a
frequency bin is the sum of the signal power and the noise power. Vaughan et
al. (1994) \nocite{Vaughan94} show that this assumption holds if a large
number, $n$, of power spectra have been averaged together. However, if $n$ is
small or $n=1$, as is the case for {\em EXOSAT} and {\em RXTE} measurements of 
4U 0142+61, this assumption is incorrect. For small values of $n$, the total
power is calculated from the vector sum of the Fourier amplitudes of the signal
and noise. Vaughan et al. (1994) \nocite{Vaughan94} estimate that upper limits
calculated using van der Klis (1989) \nocite{vanderklis89} will increase by at
least 30\% when calculated correctly. 

Upper limits for $a_{\rm x} \sin i$, the projected semi-major axis of the
neutron star, were calculated under the assumption that signal and noise
amplitudes
are combined vectorially.  No intrinsic spin frequency variations were 
detected, hence the pulse phase was given by
\begin{equation}
 \phi = \phi_0 + \nu_0 t^{\rm em}
\end{equation}
were $\phi_0$ is a constant phase and $\nu_0$ is a constant frequency. The pulse
emission time $t^{\rm em} = t^{\rm ssb}-z$ where $t^{\rm ssb}$ is the time corrected to the
solar system barycenter and $z$ is the time delay due to binary motion.  For a
circular orbit the delay is given by, 
\begin{equation}
 z = a_{\rm x} \sin i \cos \left(\frac{2 \pi (t^{\rm em} - T_{\pi/2})}{P_{\rm orb}}\right)
\end{equation}
where $P_{\rm orb}$ is the orbital period and $T_{\pi/2}$ is the epoch of
90\arcdeg\ mean orbital longitude.

The phase offsets were modeled for an arbitrary epoch $t_0$ (chosen as the
mid-time of the data set) as 
\begin{equation}
  \phi_{\rm model} = \phi_0 + \Delta \nu_0 (t-t_0) + 
A \cos \left( \frac{2 \pi (t-t_0)}{P_{\rm orb}}\right) + B \sin 
\left( \frac{2 \pi (t-t_0)}{P_{\rm orb}}\right) 
\end{equation}
for a fixed value of the trial orbital period $P_{\rm orb}$, where $\phi_0$ was a
constant offset in phase and $\Delta \nu_0$ is an offset in frequency. In 
this model, 
\begin{equation}
  a_{\rm x} \sin i = \frac{(A^2+B^2)^{1/2}}
  {\nu_0 \,\textrm{sinc}\,(\frac{\pi \Delta T}{P_{\rm orb}})}
\end{equation} 
where $\nu_0$ is the frequency used for epoch folding, 
$\textrm{sinc}\ x = \frac{\sin x}{x}$ and $\Delta T = 250$~s, the bin size. The
loss of sensitivity for short periods is accounted for by the $
 \textrm{sinc}\,(\pi \Delta T /P_{\rm orb})$ windowing factor. The phase and 
frequency offsets, $\phi_0$ and $\Delta \nu_0$, were estimated for each orbital
period, accounting for the loss of sensitivity at long orbital periods.
For each trial orbital period, single parameter 99\% confidence regions were
constructed for $A$ and $B$ (\cite{Lampton76}). Since $A = B = 0 $ was
not excluded for any trial orbital periods, an upper limit for $a_{\rm x} \sin i$
was estimated from the maximum value of $(A^2+B^2)^{1/2}$, the point on the
confidence region farthest from the origin. Upper limits on $a_{\rm x} \sin i$ at
99\% confidence generated in this manner are shown in Figure~\ref{fig:axsini}.
Sensitivity is reduced for periods $\lesssim 600$ seconds, for periods
$\gtrsim 10^5$ seconds, and at the {\em RXTE} orbital period ($\approx
96$ min). An overall 
99\% upper limit for a selected frequency range can be estimated from the
largest upper limit within that range. For all periods in the range 587~s
$\lesssim P_{\rm orb} \lesssim 2.5$ days, $a_{\rm x} \sin i \lesssim 0.27$
lt-s (99\% confidence). If we exclude the region of reduced sensitivity around
the {\em RXTE} 
orbital period, this limit reduces to 0.25~lt-s (99\% confidence). 

To search for orbital periods shorter than about 600 seconds, a different
method was employed. Well defined phase offsets and intensities could not be 
computed by cross-correlating profiles and the template for segments much 
shorter than 250 seconds. However, if the intensity was assumed to be known,
phase offsets could be computed for much shorter intervals by approximating the
cross-correlation function with a linear model. The intensity of 4U 0142+61
appeared to be reasonably constant on short time scales, so we assumed it was 
constant across 500 s intervals. Intervals 500~s long produced very good measurements of
the phase offset and intensity, while allowing longer timescale intensity 
variations. Phase offsets and relative intensities were generated for each 500
second interval by the cross-correlation method described earlier.  Next each 
500 s interval was subdivided into shorter segments, each 2 pulse periods long
($\approx$ 17.4 seconds). Fourier coefficients were then fit to each 17.4~s 
segment by minimizing 
\begin{equation}
 \chi^2 = \sum^N_{i=1} \frac{\{r_i - [c_0 + \sum^3_{k=1} u_k \cos (2 \pi k \phi
(t)) + v_k \sin (2 \pi k \phi (t))]\}^2}{\sigma^2_{r_i}}
\end{equation}
where $r_i$ is count rate measurement $i$, $N$ is the number of measurements in
a 17.4 s segment, $c_0$ is a constant offset, $k$ is the harmonic number, 
$u_k$ and $v_k$ are the Fourier coefficients, and $\phi(t) = \phi_0 +\nu_0 
(t-t_0)$ is the phase model. The Fourier coefficients then were fit by a
linearized model of the cross-correlation with $\chi^2$ given by
\begin{equation}
 \chi^2 = \sum^3_{k=1} \frac{|a_k - I_{\rm 500}(1+2 \pi i k \Delta \phi)t_k|^2}
                            {|\sigma_{a_k}|^2}
\end{equation}
where $a_k = u_k - i v_k$, $t_k$ is the template, $I_{\rm 500}$ is the intensity
for the corresponding 500~s interval, and $\Delta \phi$ is the phase
offset for each 17.4~s segment. The mean of the phase offsets for all 17.4~s
segments within a 500~s interval corresponded to the phase offset computed
for that 500~s interval using the cross-correlation method.  The top panel of 
Figure~\ref{fig:short} shows the Lomb-Scargle periodogram  generated from the 
short interval phases. The largest peak at $\approx$ 87~s has a probability of
chance coincidence of 18\%.  The bottom panel of 
Figure~\ref{fig:short} shows the 99\% confidence upper limits on $a_{\rm x} \sin i$
for fixed trial orbital periods. For the period range, 70~s $\lesssim P_{\rm
orb} \lesssim 610$~s, $a_{\rm x} \sin i \lesssim 0.26$~lt-s at 99\% confidence.

{\em EXOSAT} ME data with 1 second time resolution from 1984 August were 
obtained from the High Energy Astrophysics Science Archive Research Center 
(HEASARC). Spacecraft position information was not easily obtainable, so the 
times were corrected only for motion of the Earth. The data in 1000 second 
segments were epoch-folded at the period $P_{\rm spin} = 8.68723(4)$~s measured by 
Israel et al. (1994)\nocite{Israel94}. A template profile was generated by epoch-folding the
entire data set at the pulse period. The template and the pulse profiles from 
each segment were represented by a Fourier expansion in pulse phase. The pulse
phase offset of each segment with respect 
to the template was calculated by cross-correlating each phase fold with the 
template.  The phase offsets then were searched for an orbital periodicity 
using a Lomb-Scargle periodogram. No significant orbital signatures were found.
Using the method described earlier, upper limits were placed on $a_{\rm x} \sin
i$. For all periods in the range 2050~s $\lesssim P_{\rm orb} \lesssim 2$ days,
$a_{\rm x} \sin i \lesssim 0.73$~lt-s (99\% confidence). This limit, generated under the
assumption that signal and noise amplitudes are combined vectorially, is about
twice the limit of $a_{\rm x} \sin i \lesssim 0.37$~lt-s (99\% confidence) 
obtained by Israel et al. (1994) \nocite{Israel94}. The latter was obtained
under the assumption (\cite{vanderklis89}) that signal power and noise power
may be added, which does not apply in this case. 

\section{Discussion}   

The limits on a$_{\rm x} \sin i$, in conjunction with an assumed inclination
angle and an assumed neutron star mass of 1.4 M$_{\sun}$,
can be recast to give a limit at each orbital period on the companion mass. These 
limits include no assumptions as to the nature of the companion. A plot of
these limits, for inclination angles of 8, 30, and 90\arcdeg\, versus orbital 
period is shown in Figure~\ref{fig:mass}.

In a low-mass X-ray binary, the companion is expected to fill its Roche Lobe.
Eggleton (1983) \nocite{Eggleton83} gave a useful expression for the Roche
Lobe radius, given by
\begin{equation}
\frac{R_2}{a} = \frac{0.49 q^{2/3}}{0.6 q^{2/3} +\ln (1+q^{1/3})} 
\end{equation}
where $a$ is the separation between the two stars, $q = M_2/M_{\rm x}$ is the mass
ratio, $M_2$ is the companion mass, and $M_{\rm x}$ is the neutron star mass. This expression can be combined
with Kepler's third law to obtain the binary period as a function of the
companion star's average density, $\bar \rho = 3 M_2/4 \pi R_2^3$ and mass ratio $q$, given by
\begin{equation}
P_{\rm binary} = 3 \times 10^4\, {\rm s} \left(\frac{\bar \rho}{\bar
\rho_{\sun}}\right)^{-1/2} \left(\frac{q}{1+q}\right)^{1/2} (0.6 + q^{-2/3}
\ln(1+q^{1/3}))^{3/2} \label{eqn:periodmass}
\end{equation}
where $P_{\rm binary}$ is the binary period. If the structure of the Roche Lobe
filling star is assumed, and thus a relation between its mass and radius, the 
binary period can be expressed as a unique function of the mass, $M_2$, for an
assumed value of $M_{\rm x} = 1.4 M_{\sun}$. Period-mass relations derived 
assuming a normal main sequence star, a helium main sequence star, a white 
dwarf (\cite{Verbunt95}) and a period core mass relation for low mass
($\lesssim 2 M_{\sun}$) giants (\cite{Phinney94}) are listed in Table~2 and 
plotted in Figure~\ref{fig:mass}. The period core mass relation for low mass
giants is valid for core masses of $0.16 M_{\sun} \lesssim M_{\rm core}
\lesssim 0.45 M_{\sun}$, where the lower limit corresponds to the helium core
mass at the end of main sequence evolution and the upper limit is the core mass
at helium flash (\cite{Phinney94}). 
 
The maximum allowed mass of a Roche lobe filling hydrogen main sequence 
companion is $\sim 0.25 M_{\sun}$ for inclinations greater than 30\arcdeg\ 
(99\% confidence). This mass corresponds to an M5 or later type star. 
Larger masses are allowed if the inclination angle is quite small. For 
$i=8$\arcdeg\, the maximum mass is $\sim 0.5 M_{\sun}$ (99\% confidence). The 
probability of observing a system with $i<8$\arcdeg\ by chance is 1\%. If 
instead, a  Roche lobe filling helium burning companion is assumed, higher 
masses (99\% confidence) of $\sim 0.65 M_{\sun}$ for $i>30$\arcdeg\ and 
$2 M_{\sun}$ for $i=8$\arcdeg\, are allowed. Low mass giants, similar to the
companion to GRO J1744-28 (\cite{Finger96}), with core masses of $\sim 0.21
M_{\sun}-0.45 M_{\sun}$ are allowed for $i>30$\arcdeg\ and core masses of $\sim
0.18 M_{\sun}-0.45 M_{\sun}$ are allowed for $i>8$\arcdeg. Period-mass 
relations predict white dwarf masses well below our upper limits for all inclinations.  
 
Optical (V $\gtrsim$ 24, \cite{Steinle87}; R $\gtrsim$ 22.5, \cite{White87}) and 
infra-red (J $\gtrsim$ 19.6 and K $\gtrsim$ 16.88, \cite{Coe98}) 
observations failed to detect any counterpart within the {\em ROSAT} error circle 
(\cite{Hellier94}).  The optical limits can be used to further constrain the
allowed companion masses based upon calculated optical magnitudes of a normal
main sequence star, helium main sequence star, and the optical emission produced by 
reprocessed X-rays in the accretion disc. The column density, $N_H = 8 \times
10^{21}$ cm$^{-2}$ measured by White et al. (1996) \nocite{White96},
corresponds to an
optical extinction of $A_v \sim 4.7$ (\cite{Predehl95}), which implies a 
distance of $\sim 5$ kpc (\cite{Hakkila97}). However, the dust scattering halo
observed by {\em ROSAT} (\cite{White96}) was only half that predicted by $N_H$,
suggesting that the absorbing material is clumped along the line of sight and 
that 4U 0142+61 is most likely at a distance of $< 5$ kpc.  A main sequence 
companion with a mass of $\lesssim 0.8 M_{\sun}$ (for any inclination angle) is 
compatible with the optical limits.

The optical observations place much more stringent constraints on a giant
companion with a helium core and on a helium main sequence companion. 
The minimum core mass for a giant star with a helium core is 
$0.16 M_{\sun}$, corresponding to the core mass when the star leaves main
sequence. A giant of solar metalicity with a core mass of 0.16 $M_{\sun}$ has 
a luminosity of $L \simeq L_{\sun}$ and an effective temperature of $\sim$ 5000
K (\cite{Phinney94}). At the upper limit distance of 5 kpc with absorption $A_v
\sim 4.7$ and a bolometric correction of -0.2 (\cite{Zombeck90}), we calculate
an apparent magnitude of $V \sim 23$, for a giant star with a core mass of
$0.16 M_{\sun}$. Hence a giant companion similar to that of GRO J1744-28 is not
allowed by the optical limits. The minimum mass for a helium main sequence star
is $M \sim 0.3 M_{\sun}$ (\cite{Kippenhahn90}).  From plots in Kippenhahn \& 
Weigert (1990), a $0.4 M_{\sun}$ He star corresponds to a luminosity of 
$\sim 5.6 L_{\sun}$ and an effective temperature of $\sim$ 30,000~K.
At the upper limit distance of 5 kpc with absorption $A_v \sim 4.7$ and 
bolometric correction of -3 (\cite{Flower96}), we calculate an apparent 
magnitude $V \sim 24$ for a $0.4 M_{\sun}$ helium star. Hence a helium star of
a mass of $\lesssim 0.4 M_{\sun}$ is compatible with optical limits.

In low-mass X-ray binaries, the optical emission is usually dominated by
reprocessing of X-rays in the accretion disk.  From fits to several low-mass
X-ray binaries with known distances, van Paradijs and McClintock (1994) 
\nocite{Vanparadijs94} derived an empirical relationship between the absolute 
visual magnitude and the X-ray luminosity and orbital period. This relationship
assumes that the companion is filling its Roche-lobe, the optical emission is
dominated by the accretion disk, and that accretion disks are axially symmetric
scaled up versions of a standard shaped toy model. No assumptions are made
about the companion type. This relationship
is given by
\begin{equation}
M_{\rm v} = 1.57 - 2.27 \log \left[ \left(\frac{P}{\rm 1 hr}\right)^{2/3}
\left(\frac{L_{\rm x}}{L_{\rm Edd}}\right)^{1/2}\right]  \label{eqn:mvp}
\end{equation}
where $M_{\rm v}$ is the absolute visual luminosity, $P$ is the binary period,
$L_{\rm x}$ is the X-ray luminosity, and $L_{\rm Edd} = 2.5 \times 10^{38}$ 
erg s$^{-1}$ is the Eddington luminosity. Using this relation along with 
estimates of the X-ray luminosity, $L_{\rm x} = 7.2 \times 10^{34}$ 
d$_{\rm kpc}^2$ (\cite{White96}), absorption $A_{\rm v} \sim 4.7$ 
(\cite{White96}), and apparent visual magnitude $V \gtrsim 24$
(\cite{Steinle87}), we derived a 
relation between binary period $P$, and distance $d$. If the disk was fainter 
than the observed optical limit, then the allowed orbital periods are given by
\begin{equation}
P \lesssim 12.5 \eta\ {\rm s}\ \left(\frac{d}{1 {\rm
kpc}}\right)^{1.81} \label{eqn:pd}
\end{equation}
where $\eta$ is an estimate of the scatter in the van Paradijs \& McClintock
relationship. From Figure~2 in Van Paradijs \& McClintock (1994), we estimate 
a scatter in $M_v$ of $\sim 0.75$ magnitudes in the data used to obtain 
equation~(\ref{eqn:mvp}). A 
decrease in $M_v$ of 1.5 magnitudes (twice the scatter), results
in an increase in the maximum period by a factor of $\eta \simeq 10$.  
Thus, for the upper limit distance of 5 kpc (where $L_{\rm x} = 2 \times 
10^{36}$ erg s$^{-1}$, at which equation (\ref{eqn:mvp}) applies) this
relation, including scatter, requires a period $P \lesssim 2300$~s.
Substituting this into the period-mass relationships in Table~2, results in
allowable masses for only helium main sequence companions and white dwarf 
companions at 5 kpc. At distances $\lesssim 3$ kpc, which are more likely based
upon {\em ROSAT} scattering measurements (\cite{White96}), only white dwarf companions are 
allowed.   

\section{Conclusion}

We found no evidence (99\% confidence) for orbital modulation in the pulse
arrival times for orbital periods between 70 seconds and 2.5 days. Searches for orbital modulations
in pulse arrival times yielded an upper limit of $a_{\rm x} \sin i \lesssim $ 
0.26 lt-s (99\% confidence) for the period range 70~s to 2.5 days. Our limits 
on $a_{\rm x} \sin i$ lead to 99\% confidence dynamical limits for 
$i \gtrsim 30$\arcdeg\ of $\lesssim 0.25 M_{\sun}$ for normal main sequence 
companions, $\lesssim 0.65 M_{\sun}$ (99\% confidence) for helium main 
sequence companions, and $0.21 M_{\sun} \lesssim M_{\rm core} \lesssim 0.45 
M_{\sun}$ (99\% confidence) for $\lesssim 2 M_{\sun}$ giants with helium 
cores. Optical limits at 5 kpc allow masses of $\lesssim 0.8 
M_{\sun}$ and $\lesssim 0.4 M_{\sun}$ for normal and helium main sequence 
companions respectively. Optical limits do not allow giant companions with
helium cores. Optical and dynamical limits currently do not constrain white dwarf
companions. 

\begin{figure}
\centerline{\psfig{file=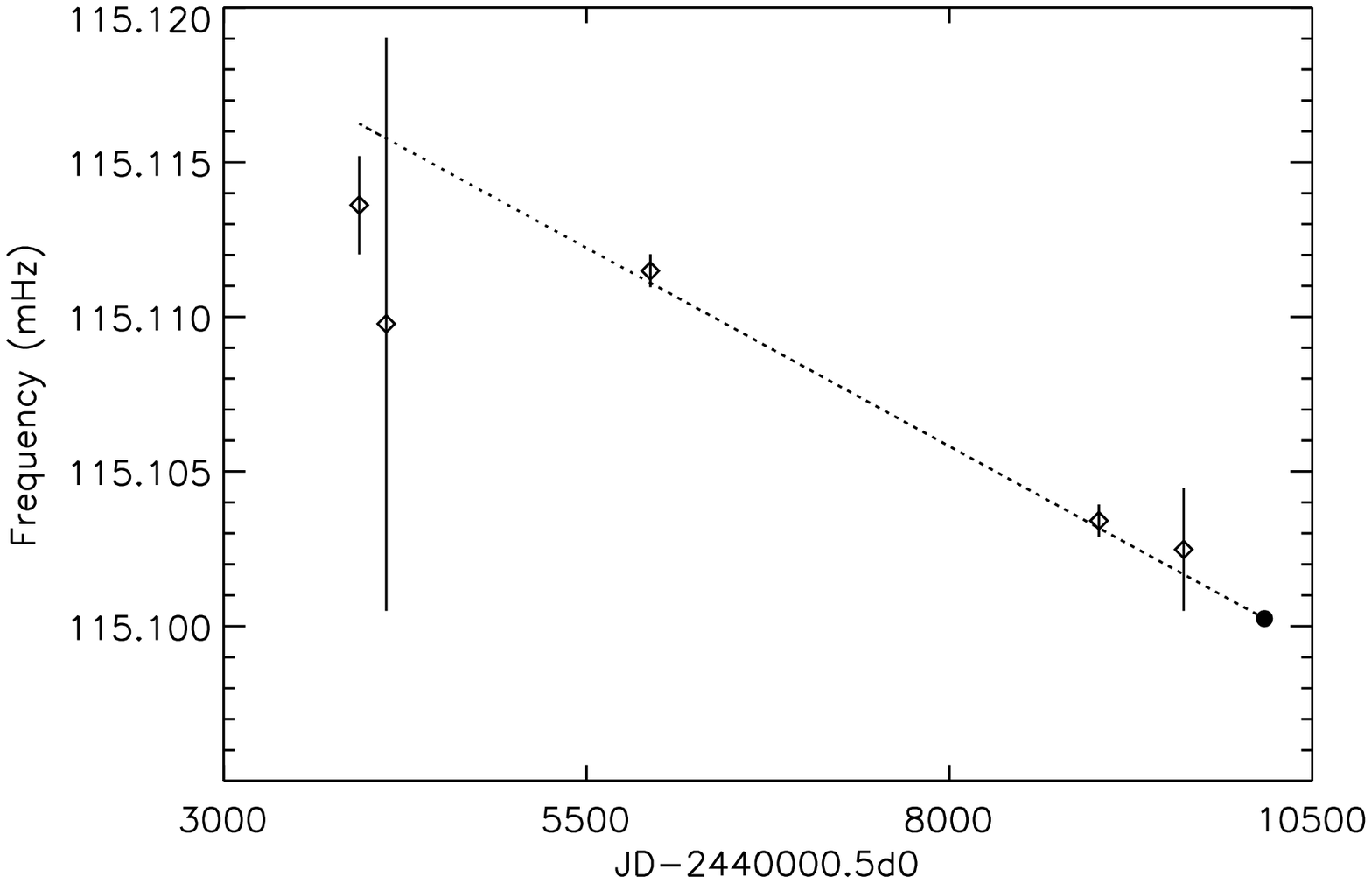,width=5in,height=3in}}
\caption{The historical pulse frequency history.  The {\em RXTE} measurement
is indicated by a filled circle. The dotted line is the best linear fit to the
overall spin down $\dot \nu = -3.1 \pm 0.1 \times 10^{-14}$ Hz s$^{-1}$.
\label{fig:period}}
\end{figure}

\begin{figure}
\centerline{\psfig{file=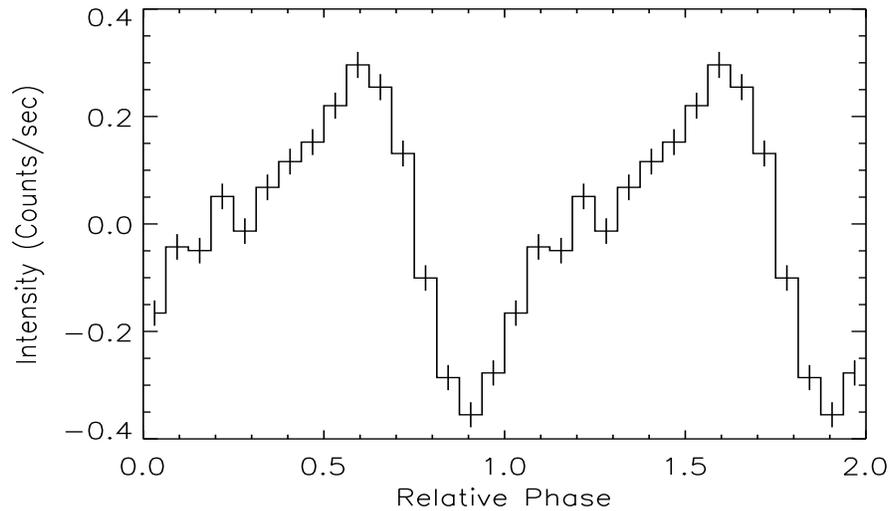,width=5in,height=3in}}
\caption{The mean-subtracted template pulse profile generated by 
epoch-folding the first 80,000 s of 3.7-9.2 keV data from March 28-30 at a 
constant frequency,
$\nu_{\rm spin}$=0.11510041 Hz.
\label{fig:profile}}
\end{figure}

\begin{figure}
\centerline{\psfig{file=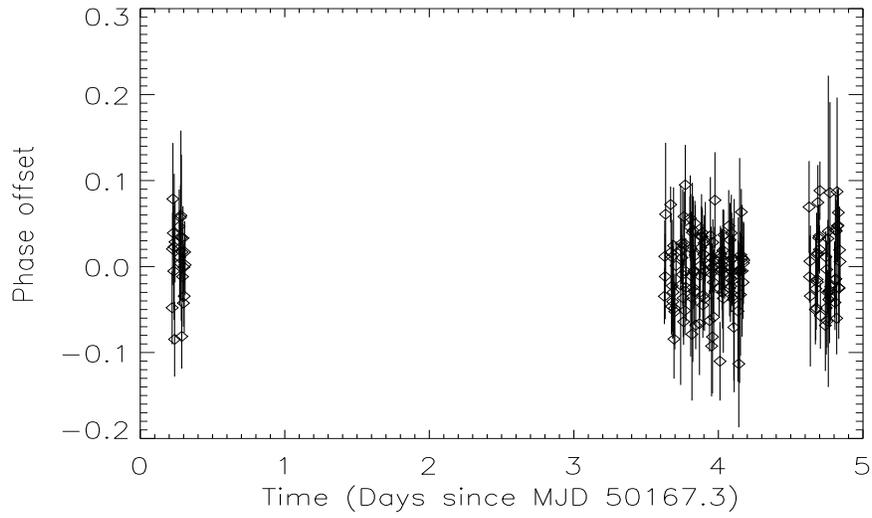,width=5in,height=3in}}
\caption{Phase offsets for 250~s intervals of data from 1996 March 25-30. A
best fit quadratic has been subtracted. The time plotted is in days since 1996
March 25:07:12:00 (MJD 50167.3). \label{fig:phases}}
\end{figure}

\begin{figure}
\centerline{\psfig{file=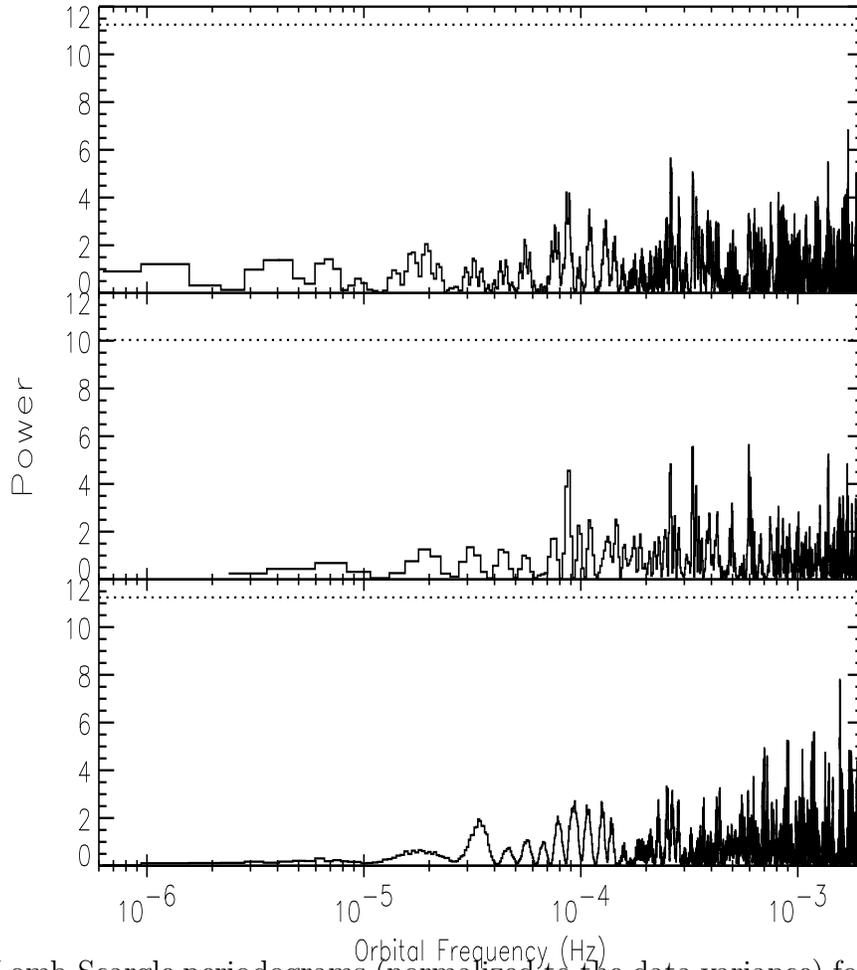,width=5in,height=5in}}
\caption{Lomb-Scargle periodograms (normalized to the data variance) for the 
entire March 25-30 data set (top
panel), the March 28-30 dataset only (center panel), and white noise (bottom
panel) with mean and variance equal to that of the phase offsets and time tags
corresponding to the March 25-30 interval.  The dotted lines indicate the 99\%
confidence level for a detection. \label{fig:lomb}}
\end{figure}

\begin{figure}
\centerline{\psfig{file=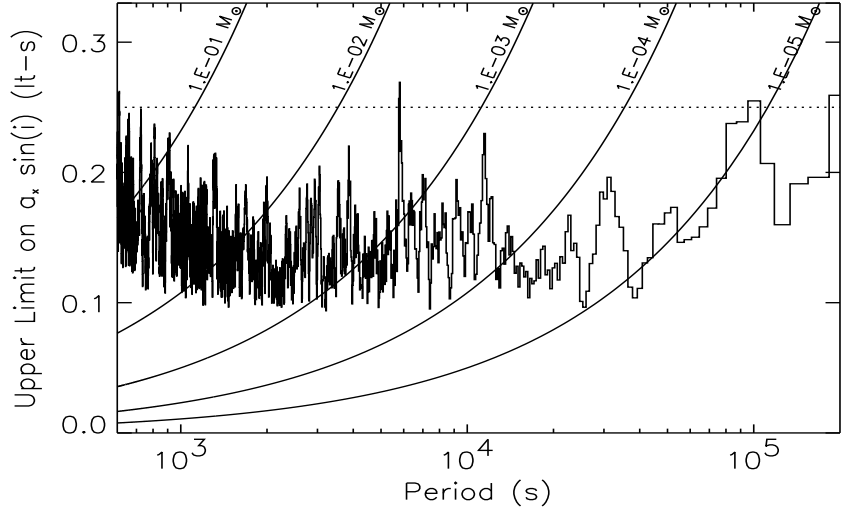,width=5in,height=3in}}
\caption{Upper limits to $a_{\rm x} \sin i$ (99\% confidence) for a circular orbit
at a trial orbital period.  Sensitivity is reduced for both the longest and 
shortest trial orbital periods. Removal of data containing Earth occultations
also reduces sensitivity at the {\em RXTE} orbital period. Curves of constant
mass function are shown. The dotted line denotes our overall 99\% confidence upper
limit of $a_{\rm_x} \sin i \lesssim 0.26$~lt-s. \label{fig:axsini}}
\end{figure}

\begin{figure}
\centerline{\psfig{file=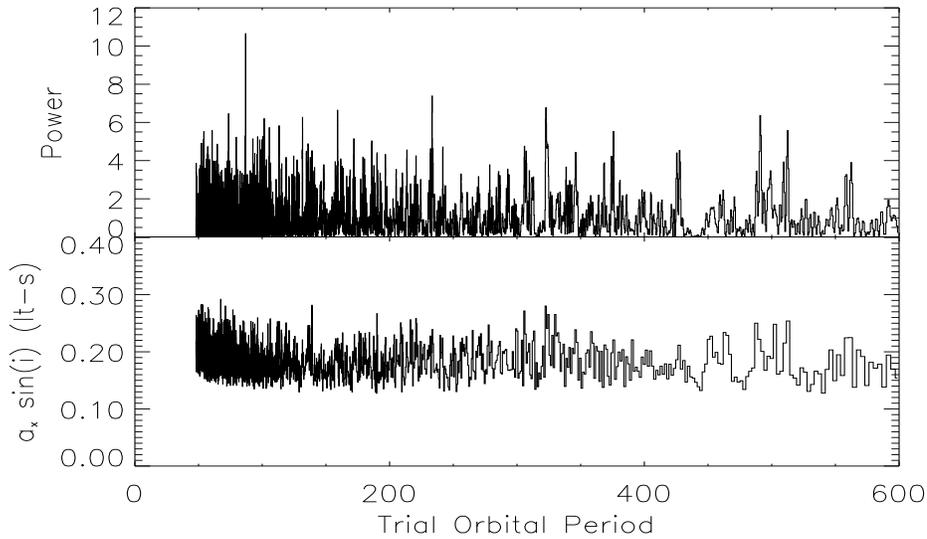,width=5in,height=3in}}
\caption{Lomb-Scargle periodogram and $a_{\rm x} \sin i$ upper limits for short ($\sim 2
P_{\rm spin}$) intervals.
\label{fig:short}}
\end{figure}

\begin{figure}
\centerline{\psfig{file=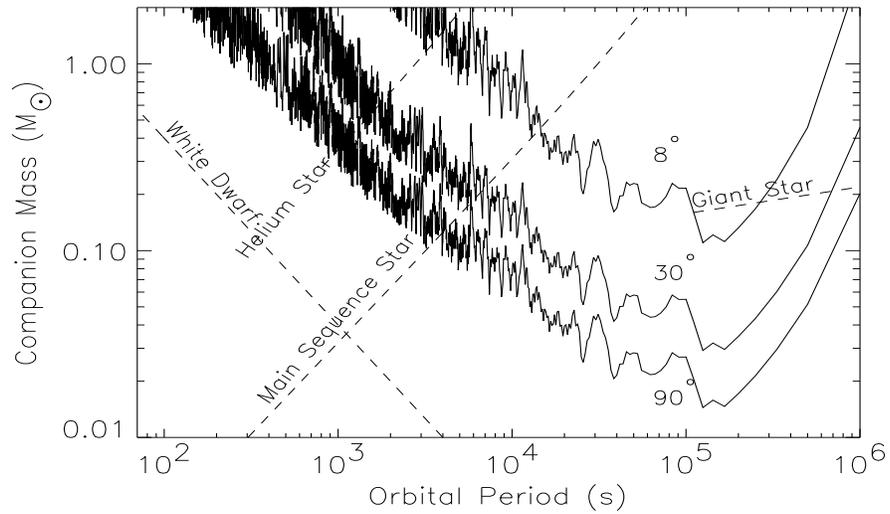,width=5in,height=3in}}
\caption{Companion mass limits calculated using the 99\% confidence limits 
on a$_{\rm x}$sin$i$ from Figures~\protect\ref{fig:axsini} \& \protect\ref{fig:short}, assuming a 1.4 
M$_{\sun}$ neutron star and 8, 30, 90\arcdeg\ inclination angles. Also shown 
are the period-mass relations for a normal main sequence star, helium burning 
star,and white dwarf and a period core mass relation for a giant star.
\label{fig:mass}}
\end{figure}

The smooth spin-down (Figure~\ref{fig:period}) observed in 4U 0142+61 is 
inconsistent with the random walk behavior expected for wind fed accreting
pulsar (\cite{Bildsten97}). Hence, 4U 0142+61 is unlikely to be a wind accretor.
Long-term observations with BATSE (\cite{Bildsten97}) show that disk fed 
accreting pulsars switch between states of spin-up and spin-down with the 
magnitude of the torque in either state comparable to a characteristic torque. 
A pulsar subject to this torque will spin up (or down) at a rate of 
(\cite{Bildsten97})
\begin{equation}
|\dot \nu| \lesssim 1.6 \times 10^{-13}\ {\rm Hz s}^{-1}\ \left(\frac{\dot M}{10^{-10} M_{\sun}\ {\rm
yr}^{-1}}\right) P_{\rm spin}^{1/3}
\end{equation}
where $\dot M$ is the mass accretion rate and $P_{\rm spin}$ is the pulsar spin
period in seconds. The observed spin-down rate of $\dot \nu = -3 \times
10^{-14}$ Hz s$^{-1}$ corresponds to $\dot M \gtrsim 9 \times 10^{-12} M_{\sun}$
yr$^{-1}$ or a luminosity of $L \gtrsim 10^{35}$ erg s$^{-1}$ which is
consistent with the luminosity of $7.2 d_{\rm kpc}^2 \times 10^{34} $ erg s
$^{-1}$ measured with {\em ASCA} (\cite{White96}). Hence the observed
long-term spin-down rate of 4U 0142+61 is consistent in behavior with the disk
fed accreting pulsars. However, optical limits only allow accretion disks in
systems with orbital periods $\lesssim 2300$~s. Orbital periods this small 
allow only helium main sequence and white dwarf companions at distances of 3-5
kpcs and only white dwarf companions at $\lesssim 3$ kpc,which are more likely based
upon {\em ROSAT} scattering measurements (\cite{White96}). However, 
current evolutionary scenarios do not provide a mechanism for producing a 
Roche lobe filling white dwarf companion or a Roche lobe filling helium main
sequence companion within the assumed $\lesssim 10^5$~yr age of the system. 
This age is based upon the low scale height of AXPs as a group 
(\cite{Vanparadijs95}) and ages inferred from supernova remnant associations of other AXPs.

Optical, dynamical, and evolutionary limits argue against binarity in AXPs. Van
Paradijs et al. (1995) \nocite{Vanparadijs95} proposed that AXPs are powered by 
accretion onto the neutron star from a circumstellar disc resulting from a
common envelope evolution. It is not clear, however, why such an evolution
would result in such a narrow range of pulse periods. The narrow period range
is much more easily explained by the magnetar model (\cite{Thompson96}). In
this model, the neutron star surface is heated by the decay of the very strong 
magnetic field, producing X-rays.  Heyl \& Hernquist
(1997) \nocite{Heyl97} also propose a model consisting of a highly
magnetized young neutron star surrounded by a thin envelope of hydrogen or
helium. In this model, X-rays are produced by thermal emission. Pulsations are
produced by a temperature gradient on the neutron star's surface induced by the
strong magnetic field combined with limb darkening in the neutron star's
atmosphere. Detailed long-term monitoring observations of
the AXPs are needed to search for spin-up episodes which would suggest the 
system is an accretor, or a glitch accompanied by a soft-gamma repeater burst
(\cite{Thompson96}), which would suggest the magnetar model. 
 
\acknowledgements{This research made use of data obtained from the High Energy
Astrophysics Science Archive Research Center (HEASARC) at the NASA Goddard
Space Flight Center. JVP acknowledges support from NASA grants NAG5-3672 and
NAG5-7105.}

\newpage

\newpage
\begin{table}[htb]
\begin{tabular}{lrc}
\multicolumn{3}{c}{Table 1: 1996 {\em RXTE} Observations of 4U 0142+61} \\
\tableline
Observation Start (TT)& Observation Stop (TT) & Observation Duration (ksec)\\
\tableline
March 25:12:31:04 & 14:14:06 & 5.533 \\
March 28:22:15:12 & March 29:00:02:06 & 3.367 \\
March 29:00:11:28 & 06:46:06 & 13.987 \\
March 29:07:23:28 & 11:34:06 & 10.182 \\
March 29:22:18:40 & March 30:03:34:06 & 11.104 \\
\end{tabular}
\end{table} 

\begin{table}[htb]
\begin{tabular}{ll}
\multicolumn{2}{c}{Table 2: Derived Mass-Orbital Period Relations for LMXBs\tablenotemark{a}} \\
\tableline
Companion Type & Mass-Orbital Period Relation\\
\tableline
Giant with He core\tablenotemark{b} & $P_{\rm orb} = 1.1 \times 10^5 $ s 
 $(M_{\rm core}/0.16 M_{\sun}) f(q)\tablenotemark{c}$\\
H Main Sequence& $P_{\rm orb} = 3 \times 10^4$ s $(M_{\rm c}/M_{\sun}) f(q) $\\
He Main Sequence& $P_{\rm orb} = 3 \times 10^3$ s $(M_{\rm c}/M_{\sun}) f(q)$\\
White Dwarf& $P_{\rm orb} = 40$ s $(M_{\sun}/M_{\rm c})\tablenotemark{d}$\\  
\end{tabular}
\tablenotetext{a}{Adapated from Verbunt and van den Heuvel (1995) \protect\nocite{Verbunt95}}        
\tablenotetext{b}{Valid for $0.16 M_{\sun} \lesssim M_{\rm core} \lesssim 0.45
M_{\sun}$ (\protect\cite{Phinney94})}
\tablenotetext{c}{$f(q) = (1+1/q)^{-1/2} (0.6+q^{-2/3} \ln (1+q^{1/3}))^{3/2}$}
\tablenotetext{d}{For $q < 0.8$, (i.e. $M_c < 1.1 M_{\sun}$), $f(q) \approx 1$} 
\end{table} 
\end{document}